\documentclass[useAMS,usenatbib]{mn2e}
\usepackage{graphicx}
\usepackage{array}
\usepackage{float}
\usepackage{hhline}

\title[]{Low-dimensional chaos in RR Lyrae models}

\author[E. Plachy et al.]{E. Plachy$^{1,2}$\thanks{E-mail:
eplachy@astro.elte.hu}, Z. Koll\'ath$^{2}$ and L. Moln\'ar$^{2}$\\
$^{1}$Department of Astronomy, E\"otv\"os University,  P\'azm\'any P\'eter s\'et\'any 1/A, H-1117, Budapest, Hungary\\
$^{2}$Konkoly Observatory, Research Centre for Astronomy and Earth Sciences, Hungarian Academy of Sciences,\\
 Konkoly Thege Mikl\'os \'ut 15-17, H-1121, Budapest, Hungary}
\begin{document}

\pagerange{\pageref{firstpage}--\pageref{lastpage}} \pubyear{2011}

\maketitle

\label{firstpage}

\begin{abstract}

The recent precise photometric observations and successes of the modelling efforts transformed our picture of the pulsation of RR Lyrae stars.
The discovery of additional frequencies and the period doubling phenomenon revealed that a significant interaction may occur between pulsational modes.
The signs of irregularities were detected both in observed light curves and hydrodynamic calculations.

In this paper we present the analysis of four peculiar hydrodynamic model solutions. All four solutions were found to be chaotic. The fractal (Lyapunov) dimensions of their attractors were calculated to be $\sim$2.2. We also investigated possible resonances between the fundamental mode and the first overtone in the dynamical neighbourhood of these models. The most important is the 6:8 resonance that was also detected in the Kepler observations of RR Lyrae itself. These results reveal that the investigation of chaotic models is important in discovering and understanding resonances in RR Lyrae stars.

\end{abstract}


\section{Introduction}
The RR Lyrae stars are known to be well-studied classical radial pulsators. According to the Bailey classification, they can pulsate either in the fundamental mode (RRab), in the first overtone (RRc), or in both at the same time (RRd). Until recently, this clear picture was complicated only by the mysterious amplitude and phase modulation called the Blazhko effect \citep{Blazhko}. The discovery of peculiar phenomena: the low-amplitude additional modes (\citealt{benko10, gug12, chadid, mosk}) and the period doubling \citep{Kepler} make this group of stars extremely interesting and a great challenge to understand. But can we expect the appearance of chaos in these variable stars?

Chaos may appear in simple deterministic systems. For certain sets of parameters the R\"ossler oscillator or a three-body system in celestial mechanics show chaotic behaviour. The evolution of chaos is nicely traceable if it emerges through period doubling. The bifurcation cascades may evolve to chaos in stellar pulsation models as well. \citet{Kovacs} presented chaotic W Virginis models. Detailed analysis showed that the dynamics are governed by two interacting vibrational modes \citep{Serre2}. The existence of chaotic pulsation was reported in several semiregular stars \citep*{Evidence}, two RV Tauri- (\citealt{R Scuti, AC Her}) and a Mira-type star \citep{Kiss} as well. These stars pulsate strongly nonadiabatically with relative growth rates of the order of ten to hundred percent \citep{10percent}. Their thermal time scale that governs phase and amplitude modulations is comparable to the length of the pulsation cycles. This explains why chaos was thought to be possible in these type of stars \citep{Serre2}. (However, chaos does not necessarily manifest in every semiregular star: variations can be driven, at least in part, by stochastic excitation ruling out the presence of low-dimensional chaos \citep{stochastic}.) Conversely, RR Lyrae and Cepheid stars show weakly nonadiabatic pulsation, their thermal and dynamical time scales differ significantly and the relative linear growth rates are small \citep{growth}. Thus chaos was not expected in classical variable stars. Although, period doubling was detected in Cepheid and BL Herculis (short-period W Vir) models (\citealt{mb90, bm92}), early RR Lyrae models did not show this phenomenon either \citep{mb90}.  

Despite of these implications the period-doubling phenomenon is observed in RR Lyrae light curves by the \textit{Kepler} \citep{Kepler} and \textit{CoRoT} space telescopes \citep{Corot} and ground-based observations \citep{Jurcsik} as well. In recent hydrodynamic calculations of the Florida-Budapest code  the period-doubling cascade was followed up to a period-eight solution \citep*{PDmodel}. The model investigations also confirmed that the destabilization is caused by the 9:2 resonance between the fundamental mode and the 9th radial overtone. Using the amplitude-equation formalism that excludes the pulsation and only calculates the amplitude variation of the modes, \citet{AE} presented not only period-doubled, but also modulated solutions arising from the 9:2 resonance.

However, beside the period-doubled ones, chaotic hydrodynamic models were also found \citep*{poster}. Further analysis showed that in these models the first overtone is also present and it is a crucial component in the occurrence of chaos \citep{Newest}. \citet{6:8} also argued that these three radial modes are present in the star RR Lyr. Detailed investigations of the three-mode RR Lyrae models will be presented in another paper (Koll\'ath et al., in prep.).

In parallel to the new findings about RR Lyrae stars, period doubling in BL Her stars \citep{smolec12} and modulation-like behaviour in BL Her models \citep{sm12} were also detected. The recent advancements in the observations and theoretical results concerning classical
pulsating variables reaffirmed our interest in the chaotic RR Lyrae model solutions.

In this paper we present the examination of four chaotic RR Lyrae model solutions of the Florida-Budapest hydrodynamic code. These models  
are displayed in Section 2. The nonlinear analysis can be found in Section 3: we briefly introduce the global flow reconstruction method and discuss the reconstruction of the radial variation and the luminosity variation of the models. 
The possible resonances that can evolve between the period-doubled fundamental mode and first overtone are examined in Section 4.  Our conclusions are summarised in Section 5.

\section{Models}

\begin{figure}
 \includegraphics[width=20pc]{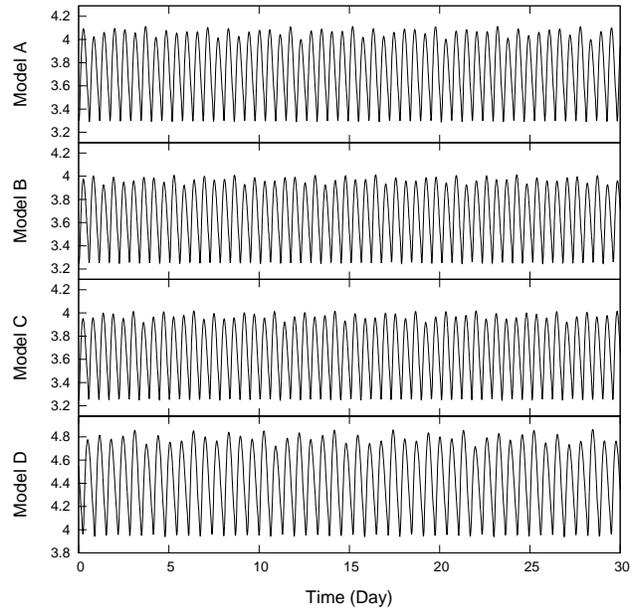}
 \caption{Radius variation of the chaotic models. Irregular variation is prominent in the maxima.}
 \label{rad var}
\end{figure}

\begin{figure}
 \includegraphics[width=20.5pc]{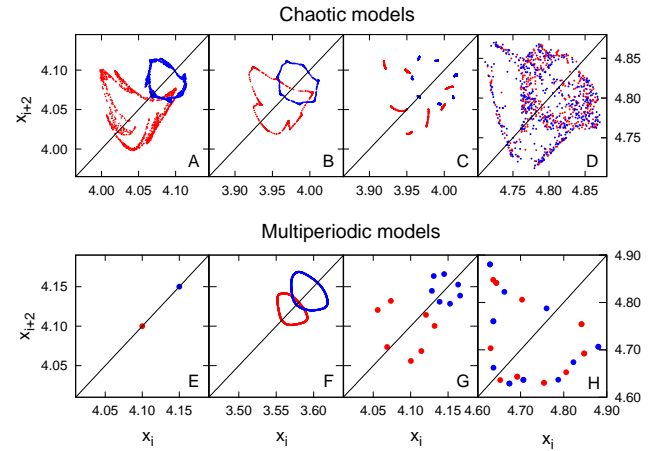}
 \caption{Upper panels: the return maps of the chaotic models that we analysed (Model A, Model B, Model C, Model D). Lower panels: the return maps of multiperiodic models: period-doubled (E), non-resonant three-mode states (F), resonant three-mode states (G) and resonant double-mode state (H).}
\label{models}
\end{figure}

We performed a dynamical investigation of four peculiar model solutions. The model parameters were: $\mathrm{T_{eff}}$ $=$ 6500~K, M $=$ 0.57 $\mathrm{M_{\odot}}$, L $=$ 40 $\mathrm{L_{\odot}}$, Z $=$ $10^{-4}$ in \textit{Model A}. The effective temperature was modified to 6360 K in \textit{Model C}. In the case of \textit{Model B} the luminosity was  also changed: L $=$ 39 $\mathrm{L_{\odot}}$. The parameters of the convective layer were similar to the ones used in the period-doubling sequences \citep{PDmodel}. These three models are the first discoveries of such peculiar behaviour in our RR Lyrae models. Parallel to this investigation the model grid has been expanded, so we have the opportunity to choose our fourth model with higher mass that is more consistent with the evolutionary predictions  of the metallicity used \citep{basti}. The parameters of \textit{Model D} are: M $=$ 0.74 $\mathrm{M_{\odot}}$, $\mathrm{T_{eff}}$ $=$ 6361~K, L $=$ 57 $\mathrm{L_{\odot}}$. For a detailed description of the code we refer to \citet{Florida1} and \citet{Florida2}.

In Figure \ref{rad var} we display the radial variation of the models. Instead of alternating high- and low-amplitude cycles, as expected from period doubling, we see an irregular variation that is especially pronounced in the maxima. 
  
The chaotic pattern is more visible in the upper panels of Figure \ref{models}, where we used the return map formalism by plotting the maxima against the second previous maximum values of the radius variation of the models. Applying this technique, the period-doubling structure can be easily recognized, as it is demonstrated in the multiperiodic models of the lower panels: the maxima alternate between the two values displaying two points in the return map (Figure \ref{models}/E). In this case the fundamental pulsation is doubled by the 9:2 resonance between the fundamental mode and the ninth overtone. 
In the non-resonant three-mode state (i.e.\ the $P_0/P_1$ period ratios are incommensurate) only two modes, the period-doubled fundamental mode and the first overtone are visible and the third mode is hidden. The system leaps between the two loops caused by period doubling and travels around them according to the relative phase of the two modes (Figure \ref{models}/F). If the three-mode state is resonant ($P_0$ and $P_1$ are commensurate) the system returns to the same phase values after a few pulsation cycles.  The pulsation modes of the model in Figure \ref{models}/G show a 14:19 resonance: the fundamental mode returns after fourteen cycles to the first phase value, during which the first overtone travels 19 cycles. \ref{models}/H displays a 20:27 resonant state without period doubling: twenty points create a single loop. This particular model will be discussed in Section \ref{dm} in more detail.

\begin{figure}

 \includegraphics[width=20.5pc]{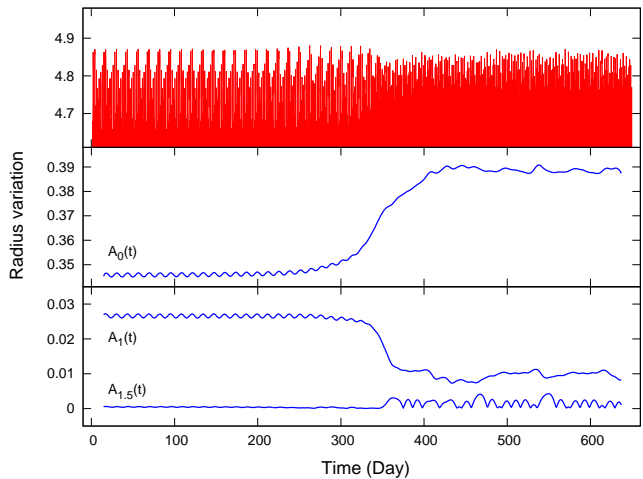}
 \caption{Time evolution of a double-mode model into chaos. Upper panel: Maxima of the radius variation. Lower panels: analitical signal functions of the amplitude of the fundamental mode ($A_0(t)$), first overtone ($A_1(t)$) and the $1.5 f_0$ subharmonic frequency ($A_{1.5}(t)$), that shows the appearance of period doubling.}
\label{timev}

\end{figure}

The return maps of the four chaotic models display a more complex structure compared to the usual quasi-one-dimensional tent or parabolic shape that is typical of chaotic systems with Lyapunov dimension of $2+\epsilon$. The return map of \textit{Model C} consists of fourteen small parabolas.
In the case of \textit{Model A} and \textit{B} we can see different distortions of the 14:19 resonance pattern. \textit{Model D} differ only by 1 K from  the resonant double-mode state displayed in \ref{models}/H. The pattern of resonance is not visible, the double-loop structure denote the presence of the third mode.

We iterated the chaotic nonlinear models up to $10^5$ cycles to rule out any transients, but the return maps remained unchanged.
Kinetic energy changes were also examined. We calculated the sum of the kinetic energy values for each pulsation cycle. Less than one percent variation was found between consecutive cycles, in agreement with typical linear growth rates in RR Lyrae models.

\subsection{A resonant double-mode solution}
\label{dm}

Beside the period-doubled and three-mode model solutions, we also found classical double-mode (RRd) models during the model survey. These models were non-resonant solutions in almost all cases, like the ones described by \citet{f0f1}. We did found, however, a very narrow range of resonant RRd models, close to the chaotic \textit{Model D} solution. Low-order resonant models have been reported in early model calculations \citep{resmod}. Here, the fundamental mode and the first overtone lock into a 20:27 resonance, corresponding to a period ratio of 0.7407. Interestingly, while the models themselves are close to the evolutionary tracks of RR Lyrae stars, the double-mode pulsation itself does not fit into the classical RRd instability strip: the period of the fundamental mode, $P_0=0.648\,d$, is higher than the theoretical upper limit of $P_0=0.62\, d$ given by \citet{f0f1}.

Although the resonant models themselves are stable, the range where they occur is so narrow that nearby models with temperature differences of only $\pm 1$ K are already chaotic three-mode models. Very little tuning of the convective parameters also returns the resonant RRd models to the chaotic state. Figure \ref{timev} shows the time evolution of a resonant double-mode model, where the eddy viscosity parameter (one of the parameters of the turbulent convection in the code) was modified slightly. The double-mode state remains stable for $~250$ days before switching to chaos. The transition itself lasts for more than hundred days. The middle and the lower panels show the evolution of the amplitude of the dominant frequency ($A_0(t)$), the first overtone ($A_1(t)$) and the $1.5 f_0$ subharmonic frequency ($A_{1.5}(t)$) respectively, calculated with the analitical signal method \citep{anfu, Florida2}. The third frequency represents the period doubling that appears during the transition from the double-mode to the three-mode state. In the same time the amplitude of fundamental mode increases and the first overtone decreases.

\section{Analyis}

\begin{figure}
 \includegraphics[width=20pc]{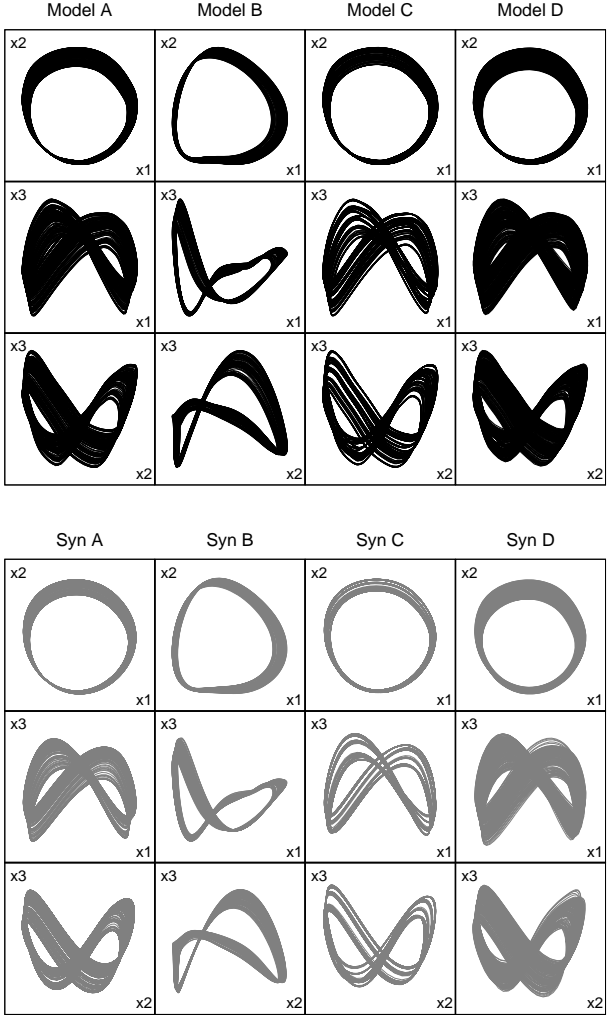}
 \caption{Broomhead-King projections of the models (black) and examples of their synthetic signals (grey).}
 \label{broom}
\end{figure}

\subsection{Method}
The global flow reconstruction technique is a nonlinear time series analyser tool suitable to detect low-dimensional chaos. Using a data sequence with equal time spacing ${s(t_n)}$ we produce the so-called delay vectors $\mathbf{X}(t_n)={(s(t_n), s(t_n-\Delta),s(t_n-2\Delta),...,s(t_n-(d_e-1)\Delta)}$, where $\Delta$ is the time delay and $d_e$ is the embedding dimension of the reconstruction space. We assume that there exists a map $\mathbf{F}$ that evolves the trajectory in time by connecting the neighbouring points, $\mathbf{X}^{n+1}=\mathbf{F}(\mathbf{X}^n)$. We fit the map $\mathbf{F}$ in polynomial form. By iterating the constructed map $\mathbf{F}$ we can produce synthetic signals that can be compared to the original data. 

The method allows us to set the time delay $\Delta$ and the embedding dimension $d_e$. The reconstruction is more successful if we broaden the phase volume of the data by adding a very small amount of noise and smoothing \citep*{Serre}. Thus we involve other adjustable parameters, the noise intensity $\xi$ and spline smoothing parameter $\sigma$. These parameters designate the parameter space where we can identify a whole region of good, robust maps. Good reconstructions provide approximate quantitative information of the system, like Lyapunov exponents and dimension. For detailed description of the method we refer to \citet{Book}.

\begin{figure}
 \includegraphics[width=20pc]{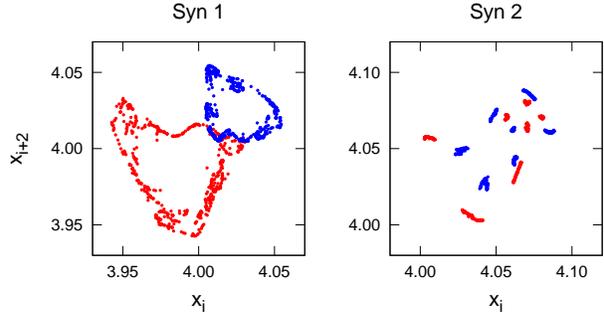}
 \caption{ The return map of \textit{Syn 1} resembles \textit{Model A} the most. \textit{Syn 2} is a reconstruction of \textit{Model C} that approaches the 14:19 resonance.}
 \label{joretmaps}
\end{figure}

In Section 4 we show a new application of the global flow reconstruction method. The synthetic signals are very similar dynamically to each other and to the hydrodynamic model solutions. Thus they give us information about the dynamic neighbourhood of the models. We use this extension of the models to investigate possible mode resonances.

\subsection{Radial variation}

We performed a global flow reconstruction of the radial variation of models in 4, 5 and 6 dimensional embedding space ($d_e$). The time delay parameter was fitted to the sampling of the data ($\Delta = 4-30$). We added gaussian noise to the test data (noise intensity, $\xi = 0 -0.0001$) and we used spline smoothing (smoothing parameter, $\sigma = 0 - 0.01$). With this relatively large parameter space we obtained several hundreds of maps that showed sufficient similarity to the original data. In Figure \ref{broom} we display the Broomhead-King projections \citep{Broomhead-King} of typical examples of synthetic signals compared to each models. 

We obtained statistically significant amount of synthetic signals to determine the quantitative properties of the original dataset. The Lyapunov dimensions are calculated to be $2.24\,\pm\,0.23$ in the case of \textit{Model A}, $2.25\,\pm\,0.23$ of \textit{Model B}, $2.17\,\pm\,0.23$ of \textit{Model C} and  $2.21\,\pm\,0.18$ of \textit{Model D}. Errors were computed as standard deviations of the Lyapunov dimension values of the ``good" synthetic signals that were generated with different parameter settings. The results of the three different embedding space ($d_e =4, 5, 6$) were handled together. 
The Lyapunov dimensions of the models are in agreement with the broad structure of the return maps. 
An independent error estimation was also carried out:  we split the $10^5$ cycles into equal fractions and reconstructed them with the same parameter combination. The standard deviation of the Lyapunov dimension values were in the same magnitude as obtained previously.

Most of the chaotic synthetic signals show the period-doubling structure.
We display the return maps of two examples of synthetic signals that resemble the models the most in Figure \ref{joretmaps}.

Beside the chaotic solutions the reconstructions provide monoperiodic and period-doubled synthetic signals, as well as period-four and -eight signals according to the the bifurcation cascade. Those synthetic signals may also display an initial stage of 50 - 100 cycles long transient chaos. 
We investigated the distance norm between chaotic and periodic maps.
We determined the quantity for this distance as $\left|\mathbf{F_1}-\mathbf{F_2}\right|=\sqrt{\frac1n\sum_{i=1}^n(F_1({x_i})-F_2({x_i}))^2}$, where $\mathbf{F_1}$, $\mathbf{F_2}$ are the maps, $n$ is the number of data points of the signal, and $x_i$ is the \textit{i}th data point. We calculated this quantity for a few randomly chosen maps. The analysis showed that there is no larger distance norm between a chaotic map and a periodic map than between two chaotic or two periodic maps. This suggests that periodic and chaotic maps are very close to each other dynamically, a small perturbation of the parameters can alter the state of the system.

\subsection{Luminosity variation}

\begin{figure}
 \includegraphics[width=20pc]{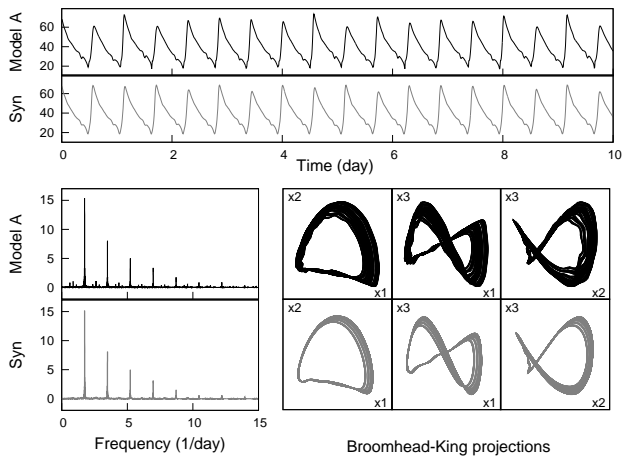}
 \caption{Time series, Fourier transformation and Broomhead-King projections of the luminosity variation (black) and its synthetic signal (grey). Values are in solar luminosity units.}
 \label{lumi}
\end{figure}

The chaos detection method needs long and quasi-continuous time series. Accuracy is also required if the amplitude variation is small. Such RR Lyrae data have been acquired recently with the \textit{Kepler} and \textit{CoRoT} space telescopes \citep{benko10, gug11}. The luminosity is a complex quantity affected by the radius and temperature changes as well as the structure of the photosphere. The reconstruction of its fine structure is challenging as the method does not reconstruct the proper frequencies, so it is expected to be less suitable for such analysis. But because there is no elaborate transformation system from single-color light curves to radial variation yet, analysis of the luminosity variation is justified and crucial if we consider a similar investigation on observational data.

We used the global flow reconstruction on the luminosity variation with the same parameters as in the reconstruction of the radius variation. However, we obtained an order of magnitude less number of successful maps. We calculated the statistical values for the Lyapunov dimension:  $2.16\,\pm\,0.12$ for \textit{Model A}, $2.05\,\pm\,0.02$ for \textit{Model B}, $2.15\,\pm\,0.13$ for \textit{Model C} and $2.12\,\pm\,0.06$ for \textit{Model D}. The values are in agreement within errors with the results of the radial variation, despite of the weaker sample. Figure \ref{lumi} displays the luminosity variation and a typical synthetic signal that we reconstructed. The latter is similar in shape, including the bump on descending branch. However, it lacks some sharp features that are more visible in the Broomhead-King projections and in the Fourier spectra. 

The complexity of the luminosity variation could not be reconstructed well. We may perform a more robust reconstruction if the luminosity variation is transformed into a filtered form. For this reason we integrated the luminosity curve after extracting the average value. As the integration is a low-pass filter we rejected the fast variations. We repeated the reconstruction method on this new dataset.
We succeeded in extending the number of good synthetic signals significantly in the case of \textit{Model A} and \textit{Model D}.
The Lyapunov dimension is calculated to be $2.20\,\pm\,0.21$ and $2.16\,\pm\,0.16$ that are in agreement with the previously obtained values. 

\begin{figure}
 \includegraphics[width=20pc]{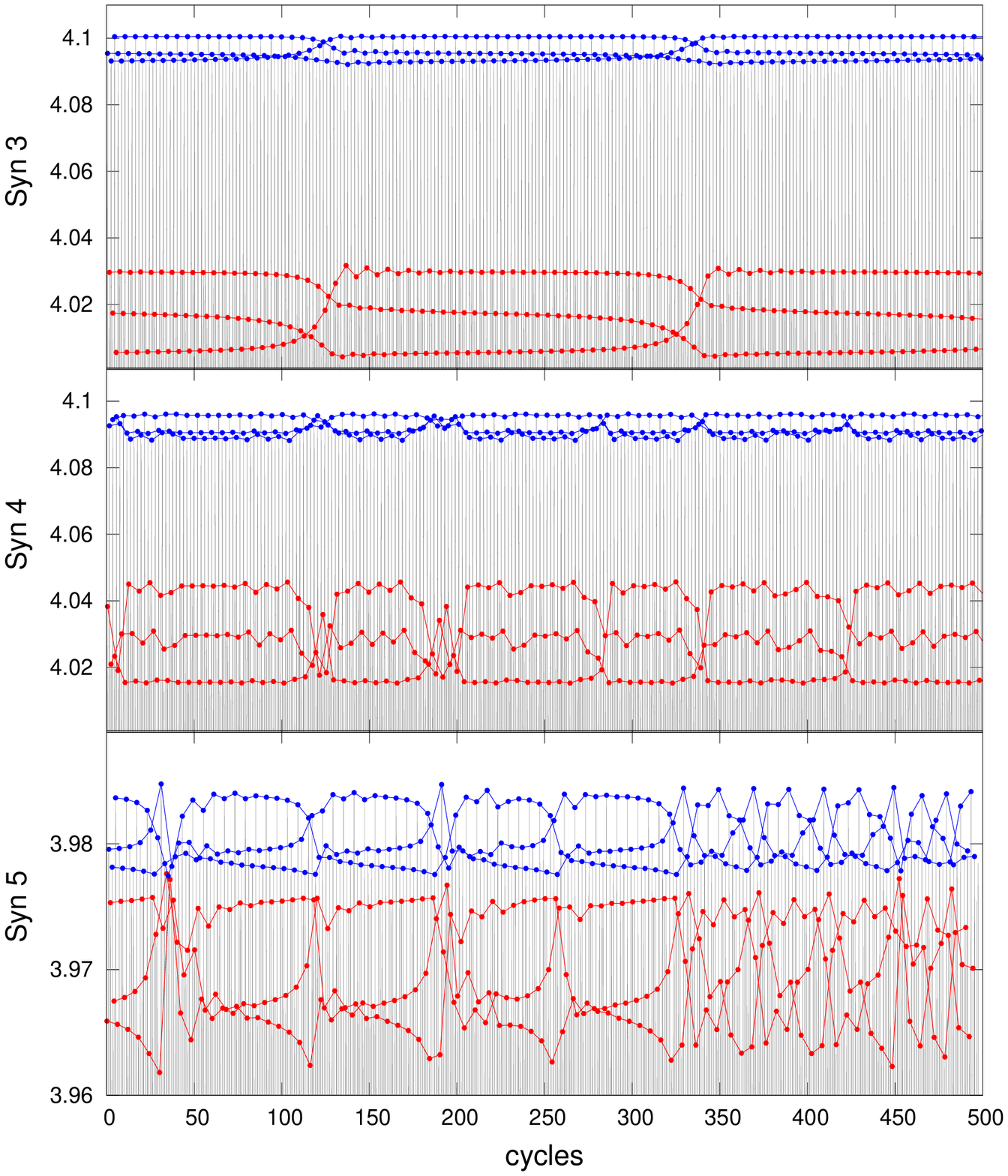}
 \caption{Chaotic synthetic signals that approach the 6:8 resonance.}
 \label{reson}
\end{figure} 
   
  \begin{figure*}
 \includegraphics[width=40pc]{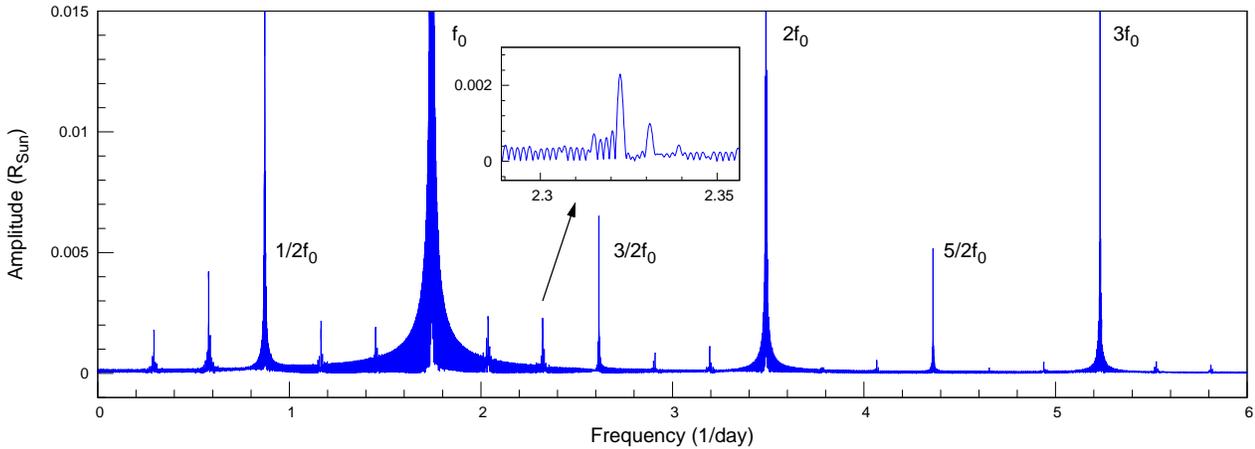}
 \caption{Fourier spectrum of \textit{Syn 3}. Harmonics ($2f_0, 3f_0,$ \dots) and subharmonics ($1/2f_0, 3/2f_0, 5/2f_0,$ \dots) are present. The lowest-amplitude peaks are caused by the 6:8 resonance, and they are split because of the irregular variations.}
 \label{ft}
\end{figure*} 

\section{Resonances}
The reconstruction resulted many maps that approach a resonant state. The resonance occurs between the bifurcated fundamental mode and the first overtone. The most frequent ratios are the 6:8 and 8:11. (We note that we use even numbers for the first element of the ratio to indicate that it is already period doubled, so we use 6:8 ratio instead of 3:4.) The 8:11 resonance can also be a period-eight solution of the bifurcation cascade. These ratios are in the accessible $\rm{P_1/P_0}$ range ($\sim$0.72-0.75) for the normal double-mode pulsation in the Petersen diagram (see Figure 6 in \citealt*{f0f1}).

Interestingly, the 14:19 resonance that is in the vicinity of the first three chaotic models does not dominate in the reconstructions. One example (\textit{Syn~2}) of the synthetic signals that approach this resonance is displayed in Figure \ref{joretmaps}. We obtained several synthetic signals that are close to a period-seven state, but lack period doubling. This periodicity was also found in the reconstruction of \textit{Model D} that contain originally a different ${P_1/P_0}$ period ratio ($\sim$20:27). This confirms that the models are also dynamically close to each other, and a reconstructed synthetic signal may resemble another model generated from different global parameters.

The 6:8 ratio is especially interesting, because it is also suspected in the \textit{Kepler} data of RR Lyrae itself (see P6 set in Figure 4 of \citealt{6:8}). In Figure \ref{reson} we show three synthetic signal examples close to the 6:8 resonance. (\textit{Syn~3} and \textit{Syn~4} are synthetic signals of the reconstruction of \textit{Model A}. \textit{Syn 5} is from the reconstruction of \textit{Model C}.) Every sixth maxima are connected. Both period-doubled maxima are tripled, and after several cycles they are reversed. Similar reversal is detected in some RR Lyrae stars \citep{Newest,6:8} and in some BL Herculis models as well \citep{sm12}. In Figure \ref{ft} we show the Fourier spectrum of \textit{Syn 3}: the peaks corresponding to the 6:8 ratio are split because of the irregular variations. We also investigated the time evolution of period-six solutions and we found that they do not necessarily evolve from bifurcated periods, they can show up after a short monoperiodic or irregular transient as well. 
We note that the 6:8 ratio was not found by the nonlinear hydrodynamic model calculations yet.

Many synthetic signal show two times five clumps in their return maps, approaching the vicinity of the 10:13 or 10:14 ratios. However, these ratios are outside of $\sim$0.72-0.75 ratio range.

Approximately half of the chaotic reconstructions do not show the period doubling of the fundamental mode, but resonances appear in those cases as well. They show most often three or five clumps in their return maps, suggesting the possibility of odd-number resonances. However, similar period-three or -five state has not been found in RR Lyrae stars yet.

The global flow reconstruction method often produces synthetic signals that deviate from the proper frequencies of the original data \citep{Evidence}. This quality was thought to be a weakness of the method until our investigations of possible resonances. Exactly these deviations make the method able to obtain a large variety of resonances providing a new application of it.

\section{Conclusion}

 We investigated four peculiar RR Lyrae hydrodynamic models with the global flow reconstruction method \citep{Book}. 
The models contain three radial modes: beside the fundamental mode, the first and ninth overtones are also present, although the
latter can only be detected through the period doubling of the fundamental mode. One of the presented models can turn to a stable double-mode state 
with a slight modification of the convective parameter. This model is the first classical double-mode solution with a high-order resonance that we found in RR Lyrae hydrodynamic calculations.
  
\begin{enumerate}
  \item We successfully reconstructed the radius variation of all four model solutions that confirms the chaotic behaviour. Although the return maps of the models show notable differences, the Lyapunov dimensions were calculated to be $\sim$2.2 in all cases.
  \item We investigated the luminosity variation of the models as well, but due to the more complex nature it turned to be less suitable for chaos detection compared to the radius variation. Nevertheless, the reconstructions are still acceptable and the computed Lyapunov dimension values are reasonable.
  \item We tested a low-pass filtered form of the luminosity variation to improve the robustness of the reconstructions. These efforts were successful for two out of four models.
  \item We found a new important usage of the global flow reconstruction method: exploration of the possible resonances. Beside the 14:19 resonance that three out of four hydrodynamic models originally approached, many other resonances and near-resonant states were found in the synthetic signals (the output of the global flow reconstruction method). 
  \item The most important resonance approached by synthetic signals is the 6:8 resonance that has been suspected in the \textit{Kepler} data of RR Lyrae \citep{6:8}, but has not been identified in hydrodynamic models yet.

 \end{enumerate}  
   
Nonlinear investigations play an important role in understanding chaotic dynamics. Irregular variations in several types of stellar pulsators originate from chaos. Here we reported an additional variable star type that is able to produce chaotic variations according to the hydrodynamic model calculations. The analysis of observational data is required to confirm this theoretical result. Our efforts showed that chaos detection is not impossible from a quantity as complex as the luminosity variation.

The first observations of \textit{Kepler} RR Lyrae stars suggested some irregularity in the period doubling. The short cadence data with one minute sampling may be ideal to decide if this irregularity really exists. Although the first results indicate that the additional variation arises from different effects (reversal of the two arms of period doubling or additional splitting, \citealt{6:8}), other stars may display true irregularity.

However, the analysis of observational data is not straightforward. The periodic and chaotic three-mode solutions are very close to each other dynamically. The return map formalism helps to distinguish between them, but difficulties may arise if the data is sampled infrequently or not uniformly. Also, almost all stars with period doubling and/or additional modes show the Blazhko effect that dominates over the patterns in the return maps and raises the embedding dimension of the reconstruction space that is limited to $d_E = 4-6$ in the method. These circumstances make the proceedings more difficult.

On the other hand, the global flow reconstruction turned out to be very helpful tool to explore the possible resonances in RR Lyrae stars. The three-mode configuration manifests itself not only in models but in stars too \citep{6:8}, and may give rise to various mode resonances. Such interactions between modes provide a possible explanation behind the Blazhko effect \citep{AE}. These investigations can help to predict the occurrence and effects of a wide range of mode resonances and---together with the exquisite Kepler data and the numerical models---may help in the development the radial mode resonance hypothesis of the Blazhko effect.

\section*{Acknowledgments}

We thank the reviewer for her/his comments that helped to improve this paper. The European Union and the European Social Fund have provided financial
support to the project under the grant agreement no. T\'AMOP-4.2.1/B-09/1/KMR-2010-0003. This work has been supported by the Hungarian OTKA grant K83790, the MB08C 81013 Mobility-grant of the MAG Zrt., the `Lend\"ulet-2009' Young Researchers' Programme of the Hungarian Academy of Sciences and the KTIA URKUT\_10-1-2011-0019 grant.

\end{document}